\documentstyle[twocolumn,aps,prl]{revtex}

\begin{document}
\draft
\title{ Self-Similarity and Localization }
\author{ Jukka A. Ketoja}
\address{ Department of Physics, {\AA}bo Akademi,\\
Porthansgatan 3, FIN-20500 {\AA}bo, Finland}
\author{ Indubala I. Satija\cite{email}}
\address{
 Department of Physics and\\
 Institute of Computational Sciences and Informatics,\\
 George Mason University,\\
 Fairfax, VA 22030}
\date{\today}
\maketitle
\begin{abstract}

The localized eigenstates of the Harper
equation exhibit universal self-similar fluctuations
once the exponentially decaying part of a wave function is factorized out.
For a fixed quantum state, we show that
the whole localized phase is characterized by a single strong coupling
fixed point of the renormalization equations.
This fixed point also describes
the generalized Harper model with
next nearest neighbor interaction below a certain threshold.
Above the threshold,
the fluctuations in the generalized Harper model
are described by a strange invariant set of the
renormalization equations.
\end{abstract}

\narrowtext
In the extensively studied Harper equation \cite{Harper}
\begin{equation}
\psi_{i+1}+\psi_{i-1}+2\lambda \cos [2 \pi (i\sigma +\phi)] \psi_i = E \psi_i,
\end{equation}
with $\sigma$ equal to the inverse golden mean,
most of the attention has
been focussed on the onset of metal-insulator transition  at $\lambda_c=1$
where the quantum states and the spectrum exhibit self-similarity and
are multifractal.
\cite{Sok}
In this paper, we show that the fluctuations in the localized
wave functions of the model for $\lambda >1$
possess the same complexity and richness
as the critical states. The universal self-similar fluctuations
at the band edges are determined by the strong coupling
fixed point of a renormalization operator.
We solve for this non-trivial
fixed point and obtain the universal
scaling ratio for the fluctuations. In addition, the stability of the
fixed point is analyzed by linearizing the renormalization
transformation. In particular, we study the pertubation associated with
a generalized Harper equation describing a two-dimensional
electron gas with next nearest neighbor (NNN)
interaction in the presence of an inverse golden mean magnetic flux (see
Eq. 12). \cite{Thou,KSC}
The universality class for the fluctuations
turns out to be unaltered as long as the NNN coupling is below a certain
threshold. Above the threshold,
the situation is lot more complicated as the
renormalization flow seems to
converge on an invariant strange set.

The wave function $\psi_i$ is written as
\begin{equation}
\psi_i = e^{ -\gamma |i|} \eta_i
\end{equation}
where $\gamma$ is the Lyapunov exponent which vanishes in the extended
(E) and the critical (C) phase. The localized  (L) phase is characterized
by a positive Lyapunov exponent corresponding to the exponential decay
of the wave function. It is assumed that the phase $\phi$ in Eq. (1)
is chosen so that the main peak of the wave function is at $i=0$ so
that $\eta_i$ is bounded.\cite{KS,Ostlund} For the Harper equation it
 has been shown analytically
that $\gamma = ln(\lambda)$ in the L phase.\cite{Aubry,Thou}
Therefore, the function $\eta_i$ describing
the fluctuations in the exponentially decaying wave function
satisfies the following tight binding
model (TBM) for $i>0$ \cite{footnote0}:
\begin{equation}
{1 \over \lambda} \eta_{i+1} + \lambda \eta_{i-1} + 2 \lambda
\cos[2 \pi (i\sigma +\phi)]
\eta_i = E \eta_i .
\end{equation}
We study this TBM using our recently developed decimation approach
\cite{JK,KS} where all sites except those labelled by the Fibonacci
numbers $F_n$ are decimated. At the $n^{th}$ decimation level,
the TBM is expressed in the form\cite{footnote1}
\begin{equation}
f_n(i) \eta(i+F_{n+1})=\eta(i+F_n) + e_n(i) \eta(i).
\end{equation}
The additive property of the Fibonacci numbers provides exact recursion
relations for the decimation functions $e_n$ and $f_n$:
\begin{eqnarray}
e_{n+1} (i)= - {A e_n (i) \over 1+Af_n (i)} \\
f_{n+1} (i)= {f_{n-1} (i+F_n) f_n(i+F_n)\over 1+Af_n(i)} \\
A = e_{n-1} (i+F_n) + f_{n-1} (i+F_n)e_n(i+F_n). \nonumber
\end{eqnarray}
In this approach, the critical phase
with self-similar wave functions is characterized by a nontrivial
asymptotic $p$-cycle (with the length $p$ equal to $3$ or $6$ for
the Harper model)
for the decimation functions.
\cite{KS} On the other hand,
the asymptotic behavior of
$e_n (0)$ and $f_n (0)$ determines the universal scaling ratios
\begin{equation}
\zeta_j = \lim_{n \rightarrow \infty} |\eta(F_{pn+j})/\eta(0)|;\;\;j=0,...,p-1.
\end{equation}

We apply the decimation method to study the fluctuations
of the L phase where
Eq. (3) provides the initial conditions for the decimation functions
$e_2 (i)$ and $f_2 (i)$ in addition to the trivial conditions
$e_1 \equiv 0$ and $f_1 \equiv 1$. As $f_2 \sim 1/\lambda <1$ in
the L phase,
the recursion relations suggest that $f_n \sim 1/\lambda^{F_{n-1}}$,
i.e. $f_n \to 0$ as $n \to \infty$. This was confirmed by the numerical
iteration of Eqs. (5) and (6).
Because $f_n$ vanishes asymptotically,
the scaling ratio $\zeta_j$ can be directly obtained as the limit of
 $|e_{pn+j}(0)|$ as $n$ tends to infinity.
At the band edges, we find that
$|e_n (0)|$ converges to
the fixed point $.1726$ for all $\lambda >1$.
Therefore, the fluctuations in the localized
wave functions are universal throughout the L phase and
are determined by the strong coupling fixed point of the system
(see Fig. 1).

Taking the limit
$\lambda \rightarrow \infty $,
the TBM reduces to
\begin{equation}
\eta_{i-1} + v \cos [ 2 \pi (i\sigma +\phi)] \eta_i = \epsilon \eta_i
\end{equation}
with $v=2$ and $\epsilon = \lim_{\lambda \rightarrow \infty}  E/\lambda$.
For the lower (upper) band edge, $\epsilon$ is
equal to $-2$ $(2)$.
At the band center,
$\epsilon =0$ and
Eq. (8) is identical to the quantum Ising model
in a quasiperiodic transverse field at the onset of
long range order.\cite{KS} Therefore, the fluctuations at the band center in
the
L phase are described by the conformal universality class of the Ising
model.\cite{KS} $|e_n (0)|$, and thus
the scaling factor
$\zeta$, exhibits asymptotically a universal
$3$-cycle $0.2307$, $0.5904$, and $0.2683$.

In order to solve for the strong coupling fixed point at the band
edges (the $3$-cycle at the band center could be analyzed in a similar way),
 we first notice that Eq. (5) can be simplified into
\begin{equation}
e_{n+1}(i) = - e_{n-1}(i+F_n) e_n(i)
\end{equation}
assuming that $f_n \equiv f_{n-1} \equiv 0$. Therefore,
this simple recursion relation describes both the asymptotic
behaviour as $n \to \infty$ for any $\lambda >1$ and the
limit $\lambda \to \infty$ where $f_n \equiv 0$ for all $n>1$.
By replacing the
discrete lattice index $i$ by the continuous renormalized
variable $x = (-\sigma)^{-n} <i\sigma>$, where $<\;\;>$ denotes
the fractional part \cite{KS}, Eq. (9) tranforms into the form
\begin{equation}
e_{n+1} (x)= -e_{n-1} (\sigma^2x+\sigma) e_n (-\sigma x) .
\end{equation}
The high $n$ limit can be studied
by introducing the renormalization operator
\begin{equation}
T [u(x),t(x)] = [t(-\sigma x), -u(-\sigma x -1) t(-\sigma x)] .
\end{equation}
The importance of this operator follows from the fact that
the recursion (10) can be represented as the operation by $T$
on the pair $[e_{n-1} (-\sigma x), e_n (x)]$.
Moreover, the limiting function $e^*(x)=\lim_{n\to \infty} e_n(x)$
 corresponds to
the fixed point $[e^*(-\sigma x),e^*(x)]$ of the operator $T$.
An initial estimate of the fixed point can be obtained by
applying $T$ subsequently on the pair $[e_2 (-\sigma x), e_3(x)]$
obtained from Eq. (8).
It turns out that it is easier
to expand $1/e^*(x)$ than $e^*(x)$ (they both satisfy
the same fixed point equation)
so we solve for the coefficients of $1/e^*(x)$ by truncating
the series and applying the Newton method to determine the fixed point
of $T$. The power series is convergent in the domain
$|x| \leq 1$, and we can obtain better and better estimates for the
principal scaling ratio $\zeta =|e^*(0)|$ by increasing the order of the
power series. Including terms upto the order $x^{23}$ we observe that
$e^*(0)$ approaches $-0.172586410945 \pm 10^{-12}$
in agreement with the result obtained by iterating
the decimation equations.\cite{footnote3}

The linear stability analysis at the fixed point shows
that the renormalization operator $T$
 has the unstable eigenvalues $\pm \sigma^{-2}$
and $\pm \sigma^{-1}$ ($\sigma^{-1}$ is a double eigenvalue)
and the marginal eigenvalue $-1$.
In addition, there is
a set of stable eigenvalues which are powers of the inverse golden mean.
It should be noted that our renormalization
operator resembles the one of Ostlund and
Pandit \cite{Ostlund} for the study of the critical point of the
Harper equation. Although, in their case $t$ and $u$ are $2\times 2$
matrices, the eigenvalue analysis is similar in their and our case.
The variation of $\epsilon$ or $v$ in Eq. (8) leads to an
asymptotic escape from
the fixed point in the eigendirection associated with
the unstable eigenvalue $\sigma^{-2}$. In the same way,
the variation of $\phi$ can be related to the eigenvalue
$-\sigma^{-1}$. We expect that some
 of the remaining unstable eigenvalues
and the marginal one
are not "physical" because they represent variations which are
inaccessible
by using a TBM to define the pair $[u,t]$ (see ref. \cite{Ostlund}).
We could further generalize the analysis by considering the direction
of a finite decimation function $f_n$ in function space but as explained
previously, the pertubation in that direction is expected to be
irrelevant (with eigenvalue $0$).

It is a characteristic feature of
the strong coupling fixed point that
all eigenvalues are powers
of the golden mean. In the critical case, where the spectrum is
singular continuous, there is a non-trivial eigenvalue associated
with change of energy. We do not expect such an eigenvalue to
appear in the localized phase where the spectrum is point like.
In fact, the appearence of the eigenvalue $\sigma^{-2}$ can be traced to
the dual system \cite{Aubry} for which $\lambda \to
\infty $ is the weak coupling limit. \cite{Ostlund}

We next study the generalized Harper equation
\begin{eqnarray}
\{ 1+\alpha \cos[2 \pi (\sigma(i+\frac{1}{2})+\phi)]\} \psi_{i+1}\nonumber\\
+ \{ 1+\alpha \cos[2 \pi (\sigma(i-\frac{1}{2})+\phi)] \} \psi_{i-1}
\nonumber\\
+ 2 \lambda \cos[2 \pi (\sigma i+\phi)] \psi_i
=E\psi_i
\end{eqnarray}
describing Bloch electrons on a square
lattice with nearest neighbor (NN) coupling anisotropy $\lambda$ and NNN
coupling $\alpha$. This model was recently
studied using analytical and numerical methods\cite{Thou} as well as
applying
the decimation scheme.
\cite{KSC} The model was found to exhibit a fat C phase provided $ \alpha \geq
1$ and $\lambda \leq \alpha $. In the fat C phase the model exhibited various
new
universality classes:  at certain specific values of $\lambda$ and
$\alpha$, the self-similarity of the critical wave functions
was described by higher order unstable (with respect to a change of
parameter values) limit
cycles
 of the RG equations  while for arbitrary values of the parameters, the
RG flow converged on an invariant set.
This implied that the fractal characteristics of the wave functions were
not self-similar.
Since the cycle lengths
were as high as $24$, the question of which parameter
values exhibited a limit cycle and which converged on strange set remained
somewhat open.

We now apply the procedure outlined above to study the fluctuations in the
L phase of the generalized Harper equation (12).
We again make use of the explicit formulas of the
Lyapunov exponents as obtained from ref. \cite{Thou}.
Numerical iteration of the decimation equations shows that
for $0 \leq \alpha < 1$, the fluctuations in the L phase are
determined
by the same strong coupling fixed point as for the Harper equation.
 For $\alpha
\geq 1$, the decimation functions for the fluctuations in the L phase
are found to flow away from the the strong coupling fixed point.
However, in analogy with the Harper case,
the asymptotic behavior of the decimation
functions appears to be independent of the value of $\lambda$ throughout
the L phase and is described by the same renormalization equations
(10) and (11). That is, for all values of $\alpha$, the decimation
function $f_n$ renormalizes to zero.
Thus we can focus on the limit
$\lambda \rightarrow \infty$, where $E/\lambda$ tends to
$\pm 2$ for the band edges, and the generalized TBM for $\alpha \geq 1$
reduces to
($\phi$ is $1/2$ for the negative and $0$ for the positive band edge)
\begin{equation}
\{\pm 1+\alpha \cos[2 \pi \sigma(i-{1\over 2})] \} \eta_{i-1} =
\alpha [1-\cos(2 \pi \sigma i )] \eta_i
\end{equation}
Because of this simple TBM form (containing no parameter such as the energy
which would be known to limited precision), the recursion relation (9) for the
decimation function $e_n$ can be iterated upto $35$ times
thereby studying systems of the size $14930351$.
This accuracy is
particularly crucial in order to observe
higher order limit cycles.
Our detailed numerical study reveals
various limit cycles at certain values of $\alpha$.
Writing
$\alpha=1/abs(cos( 2 \pi r))$, the limit cycles are observed for the rational
values of $r$.
For example, for $r=1$, the period $p$ of the limit cycle
is found to be $3$, while $p=12$
for the values $r=1/3, 1/6, 1/8, 1/14, 1/18, 2/9$. Furthermore, we observe
$p=24$ for $r=1/7, 1/12, 1/16,3/14, 3/16, 3/23$
and $p=18$ for $r=1/17, 1/19, 3/17, 3/19$.
Since the RG equations cannot be iterated more than $35$ times (because
of memory limitations),
we cannot see higher order cycles.
Based on this study, we conjecture that for all rational values of $r$,
the RG flow converges on a limit cycle of period which is a multiple of $3$.
However, the correlation between $r$ and $p$ still remains a mystery.

For arbitrary values of $\alpha$, the decimation functions
do not converge on a limit cycle. The plot of $e_n(0)$ vs
$e_{n+1}(0)$ is found to converge on an invariant set which resemples
an $orchid$  flower (see Fig. 2). We conjecture that the invariant set
of
the renormalization operator is
a universal strange attractor.
The periodic orbits corresponding
to the rational values of $r$ are expected to be dense on the attractor.
However, we conjecture that the probability to hit a periodic orbit
is still zero.
The confirmation of these ideas by an explicit solution of
the renormalization  equations
remains open.

It turns out that
the results at the strong coupling limit of the generalized Harper equation
shed some light on the C phase of the model.
\cite{ft2}
This is due to the fact that for a fixed value of $\alpha$,
the period of a strong coupling limit cycle
coincides with the period of a similar
limit cycle at the $critical$ line $\lambda =\alpha$.
Analogously, the existence of a strange
set in the strong coupling limit strengthens our previous conjecture\cite{KSC}
on the existence of a similar set in
the C phase.

In summary, our decimation studies show the existence of a new strong
coupling renormalization fixed point which controls the universal
fluctuations of the localized wave functions in the Harper equation.
Unlike the trivial
fixed point of the weak coupling limit, the strong coupling fixed point
describes a new non-trivial universality class for the Harper equation.
In analogy with the critical phase, these fluctuations
are characterized by a universal scaling
ratio
which is determined by the value of the fixed point function at the origin.
We are able to find a power series expansion of the
fixed point and examine the stability of the fixed point under the
renormalization. It turns out that a change in  the NN or NNN coupling
is an irrelevant perturbation as long as the NNN coupling is below
a certain threshold. Above the threshold, the system is outside
the basin of the attraction of the fixed point and
the renormalization flow is attracted by an invariant strange
set of the function space.
Our studies put the localized and
the critical phase on the same footing. The localized phase could in
fact be viewed as a fat critical phase.

In some recent studies of the Anderson problem, a
controversy persists concerning
the multifractal character of the non-exponential part of the wave
functions.
\cite{PSTZ,SG,Lema} The major problem in various numerical studies
is that due to limited precision,
it is not clear whether the studies really describe the properties
of an infinite system. The Harper equation and its generalization provide
the simplest class of one-dimensional models exhibiting localization
where the Lyapunov exponent is known analytically. Therefore,
these models contain no parameter which limits the precision of
iteration of the decimation functions so that extremely
large systems can be studied.
Our results for the self-similar
fluctuations in localized states of quasiperiodic
systems strengthen the likelyhood of fractal characteristics in
random or aperiodic systems exhibiting localization.
Since localization is generic
in a variety of systems\cite{loc} including the models of
quantum chaos\cite{QC}, the results described here open a new avenue
in the localization theory.

\acknowledgements

The research of IIS is supported by a grant from National Science
Foundation DMR~093296. JAK is grateful for the hospitality
of the Research Institute for Theoretical Physics in Helsinki.

\begin{figure}
\caption{ The universal fluctuations of a band-edge eigenstate in
the L phase of the Harper equation. Note that unlike the critical
phase wave function, the L phase fluctuations are not symmetric about the
secondary peaks. }
\label{fig1}
\end{figure}

\begin{figure}
\caption{ A two-dimensional projection  of the attractor of the renormalization
 flow
 in the strong coupling
limit $\lambda \to \infty$. For about $4000$ different values of $\alpha >1$,
decimation equations were iterated $35$ times and the first $10$ iterates
were ignored as transients.}
\label{fig2}
\end{figure}


\begin{references}
\bibitem{footnote0}
The fluctuations for $i<0$ satisfy the same TBM except that
$\eta_{i+1}$ is multiplied by $\lambda$ and $\eta_{i-1}$ by $1/\lambda$.
This causes the fluctuations to be symmetric around $i=0$.
\bibitem{footnote1}
The decimation equation here is written in a slighly
different form from our previous papers\cite{KS} in order to have
bounded asymptotic limits of the decimation functions $e_n$ and $f_n$ as
$n \to \infty$ also in the localized case $\lambda >1$.
\bibitem{footnote3}
If the absolute value is not taken, the fixed point is observed
only at the positive band edge. The negative band edge corresponds
to the $3$-cycle $[-e^* (-\sigma x),e^*(x)]$,
$[e^* (-\sigma x),-e^*(x)]$, and $[-e^* (-\sigma x),-e^*(x)]$.
\bibitem{ft2}
Due to limited precision in the energy, the decimation equations
can be iterated only about $28$ times in the C phase. However, in the strong
coupling limit it is possible to go upto $35$ iterations.
\end{references}
\end{document}